\documentclass[
reprint,
raggedbottom,
groupedaddress,
showpacs,
 amsmath,amssymb,
 aps,
 prl,
 showpacs
]{revtex4-1}

\usepackage{graphicx}
\usepackage{dcolumn}
\usepackage{bm}
\usepackage{xcolor}
\usepackage{soul}
\usepackage[utf8]{inputenc}
\usepackage{comment}

\begin{document}

\preprint{APS/123-QED}

\title{Morphological evolution of an Au crystalline domain in a nano-particle investigated by Bragg coherent diffraction imaging}
\author{Seonghyun Han}
\affiliation{Department of Physics and Photon Science, Gwangju Institute of Science and Technology, 61005 Gwangju, Korea
 }
\author{Saehyun Kang}
\affiliation{Department of Physics and Photon Science, Gwangju Institute of Science and Technology, 61005 Gwangju, Korea
 }
\author{Ouyoung Kwon}
\affiliation{Department of Physics and Photon Science, Gwangju Institute of Science and Technology, 61005 Gwangju, Korea
 } 
 \author{Su Yong Lee}
 \affiliation{Pohang Accelerator Laboratory, Pohang, 37673, Republic of Korea
 }
\author{Wonsuk Cha}
\affiliation{Advanced Photon Source, Argonne
National Laboratory, Lemont, IL 60439, USA
 }
\author{Hyon Chol Kang}
\email{kanghc@chosun.ac.kr}
\affiliation{Department of Materials Science and Engineering, Chosun University, 61452 Gwangju, Korea
 }
\author{Chan Kim}
\email{chan.kim@xfel.eu}
\affiliation{European XFEL, 22869 Schenefeld, Germany}
\author{Do Young Noh}
\email{dynoh@gist.ac.kr}
\affiliation{Department of Physics and Photon Science, Gwangju Institute of Science and Technology, 61005 Gwangju, Korea
 }%

\date{\today}

\begin{abstract}
Understanding the dynamic structural evolution of metallic nanoparticles is crucial for tailoring their functional properties. In this study, we utilized three-dimensional coherent X-ray diffraction to investigate the morphological transformation of Au nanocrystalline domains in a multi-domain nanoparticle during \emph{in-situ} annealing. 
By combining Bragg coherent diffraction imaging with autocorrelation function analysis, we observed a transformation from an anisotropic multi-domain configuration toward a single crystalline equilibrium morphology.
Above 600~$^{\circ}$C, a small subsidiary domain merged into the primary domain, which was continuously coarsened at 625~$^{\circ}$C.
Finally, at 635~$^{\circ}$C, the low-energy $\{111\}$ and $\{100\}$ facets emerged, and the crystal shape approached the equilibrium truncated octahedral geometry.
By elucidating complex domain dynamics, this work offers fundamental insights into domain growth and equilibrium crystal shape formation.
\end{abstract}

\maketitle

Gold nanoparticles (AuNPs) have played a pivotal role in various catalytic reactions due to their remarkable stability, high activity in specific chemical reactions, and efficiency at low temperatures \cite{ishida2020,haruta1997,hutchings2005}.
These unique properties have led to widespread applications in biomedical, imaging, sensing, and optical technologies \cite{dykman2012,saha2012,si2021}.
The physicochemical properties of AuNPs sensitively depend on their size, shape, and crystallinity, enabling precise modulation of reactivity \cite{haruta1997,daniel2004,liang2022,carone2023}.
Nanoparticles (NPs) with multi-domain structures contain internal defects and domain boundaries, which may alter charge transfer and reactive sites \cite{pacchioni2018,zachman2022}. In contrast, single-domain single crystalline particles exhibit enhanced physicochemical properties in catalytic reactions due to their uniform electronic structure and stable crystallographic properties \cite{hu2025,lim2022}. Therefore, understanding the transformation from multi-domain to single domain and the growth of related crystallographic planes during this process is crucial for controlling the functional properties of NPs.
In addition, there has long been fundamental interest in understanding the kinetics of crystalline domain growth and how it relates to their equilibrium crystal shape \cite{rheinheimer2020,elbaum1993,wang2014,yang2022,dzhigaev2022}, particularly under catalytic reaction conditions \cite{bachmann2025}.

While the morphological evolution of individual AuNPs has been extensively studied, understanding the growth of single-crystalline domain within a polycrystalline nanoparticle remains a significant challenge. 
The challenge stems from the difficulty in observing a growing internal crystal domain in the presence of surrounding domains with a sufficient spatial resolution.
Conventional diffraction techniques and electron microscopy alone are limited in directly observing this structural evolution of an individual crystalline domain.
Coherent diffraction imaging in Bragg geometry \cite{robinson2009,mastropietro2017,kim2018} is unique 
in its ability to non-destructively probe the three-dimensional (3D) structure and internal strain distribution in nano-particle crystal domains, enabling \textit{in-situ} tracking of their dynamic changes under various conditions \cite{yau2017,vicente2021,park2024}.

Coherent X-ray diffraction intensity measured in a finite reciprocal space volume $V_Q$ surrounding a Bragg peak $\mathbf{G}$ can be described: 
\begin{align}
   I(\mathbf{q}) \sim 
     \left |\int  d\mathbf{R} ~ \rho_\mathbf{G}^c (\mathbf{R}) \exp ( i \mathbf{q} \cdot \mathbf{R})  \right |^{2} ,
\end{align}
where 
$\mathbf{q}$ is the momentum transfer relative to $\mathbf{G}$, and
$\rho_{\mathbf{G}}^c (\mathbf{R}) \left(= 
  \rho_{\mathbf{G}}(\mathbf{R})e^{i\phi (\mathbf{R})} \right) $ is a complex local order parameter characterizing the lattice order of a NP, which is 
the Fourier component of the atomic electron density $\rho(\mathbf{r})$ at $\mathbf{G}$, coarse-grained over volume V.
V is about the reciprocal of $V_Q$, i.e, $(2\pi)^3/V_Q$ centered at $\mathbf{R}$, and supposedly much larger than an atomic unit cell volume.
In Bragg coherent diffraction imaging (BCDI), $\rho_{\mathbf{G}}(\mathbf{R})$ and $\phi (\mathbf{R})$ are reconstructed using the amplitude of its Fourier transform measured experimentally, $\sqrt{I_q}$, and the phase retrieved using various phase-retrieval algorithms \cite{fienup1982,robinson2009,miao2012,minkevich2008}.
BCDI can track crystalline domain and strain evolution in response to sample environments non-destructively \cite{robinson2009, ulvestad2016,dzhigaev2022}, making it an ideal probe for investigating structural changes \textit{in-situ}.

In addition, the direct Fourier transform of the diffraction intensity, $I (\mathbf{q}) $ provides the autocorrelation function (ACF) of order parameter:
\begin{align*}
   ACF(\mathbf{R}) \sim 
     \int  d\mathbf{R'} ~ \rho_\mathbf{G}^c (\mathbf{R'})\rho_\mathbf{G}^{c*} (\mathbf{R'}-\mathbf{R}) .
\end{align*}

The ACF derived from 3D coherent X-ray diffraction intensity provides a robust methodology for characterizing particle evolution, as it is inherently free of various image reconstruction artifacts \cite{patterson1934,robinson2009}.
Unlike the ensemble-averaged autocorrelation function (ACF) derived from incoherent X-ray diffraction, the coherent X-ray ACF resolves structural correlations within a specific illuminated area.
Particularly for specimens undergoing structural transitions during measurement, the ACF offers a meaningful temporal average.
While the ACF can serve as an effective initial density estimate for BCDI, its greater utility lies in its capacity to illustrate overall shape of an object qualitatively.

In this study, we utilized \textit{in-situ} Bragg coherent diffraction imaging and coherent X-ray ACF to investigate the morphological evolution of a crystalline domain in a multi-domain AuNP during annealing.
Morphological changes—including coarsening, and faceting—were observed as the initially small and irregular crystalline domain grew toward its equilibrium shape.

\begin{figure}[t]
    \centering
    \includegraphics[width=1\linewidth,trim=0cm 3.5cm 0cm 0cm,clip]{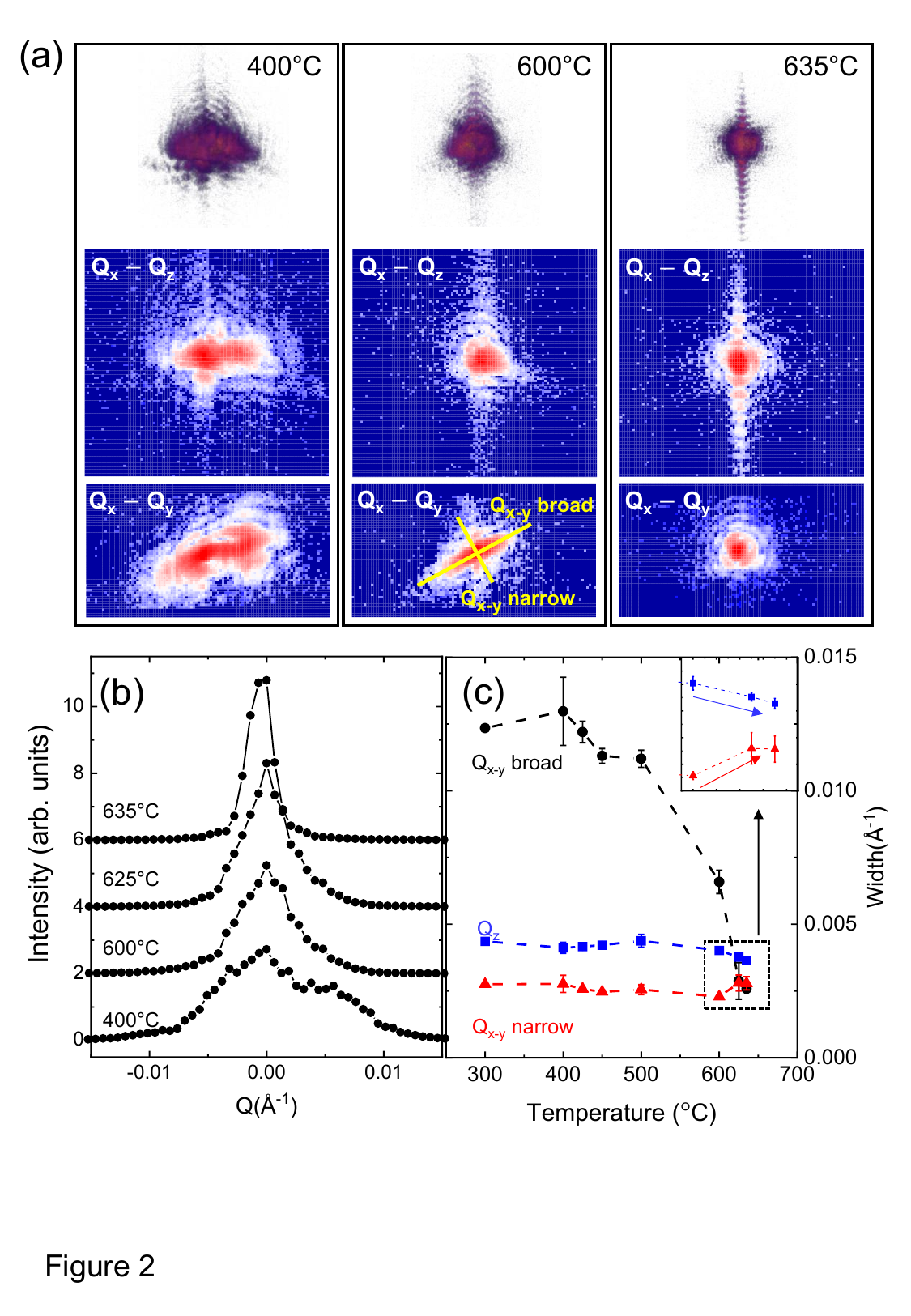}
    \caption{(a) 3D coherent diffraction profiles around the Au(111) Bragg peak at 400~$^\circ$C, 600~$^\circ$C, and 635~$^\circ$C, showing the \mbox{$Q_\mathrm{x}$--$Q_\mathrm{z}$}  and \mbox{$Q_\mathrm{x}$--$Q_\mathrm{y}$}  cross-sections. At 635~$^\circ$C, the fringe patterns become prominent. 
    (b) Line profiles along the in-plane direction ($Q_{x\text{–}y}$ broad direction).
    A double-peak structure is shown at 400~$^\circ$C,
    which transitions to a nearly single-peak profile upon heating above 600~$^\circ$C.
    (c) Widths of the XRD profile in the $Q_{z}$ (blue), $Q_{x\text{–}y}$ broad (black), and $Q_{x\text{–}y}$ narrow (red) directions.}
    \label{fig:fig2}
\end{figure}

As the annealing temperature increases, the Au(111) Bragg peak becomes sharper and better-defined. 
Figure~\ref{fig:fig2}(a) presents the 3D diffraction patterns stabilized at 400~$^\circ$C, 600~$^\circ$C, and 635~$^\circ$C, alongside the \mbox{$Q_\mathrm{x}$--$Q_\mathrm{z}$} and \mbox{$Q_\mathrm{x}$--$Q_\mathrm{y}$} slices, where the $Q_z$ direction is parallel to the Au $<111>$ direction.
Below  400~$^\circ$C, the scattering intensity is broad and anisotropically distributed in the \mbox{$Q_\mathrm{x}$--$Q_\mathrm{y}$} plane, 
indicating that the initial crystalline domains contributing to the Au(111) reflection are small and anisotropic in real space. 
The diffraction signal in the $Q_x$ direction consists of two broad peaks at this temperature as shown in the line profiles in Fig.~\ref{fig:fig2}(b).
This suggests that, among multiple domains, two specific crystalline domains with slightly different crystallographic orientations satisfying the Bragg conditions within the coherent volume, are primary contributors to the diffraction intensity.
The peaks merged into a single-peak at 600~$^\circ$C which was still elongated in the \mbox{$Q_\mathrm{x}$--$Q_\mathrm{y}$} plane along the long axis direction indicated in the second panel of Fig. \ref{fig:fig2}(a).
Upon heating to 635~$^\circ$C, however, the peak sharpened and became more isotropic, reflecting crystal domain growth and morphological change.
Furthermore, the speckle patterns become more pronounced developing crystal truncation rods along a few specific crystallographic directions-indicating the formation of distinct facets during annealing.

\begin{figure}[!t]
    \centering    \includegraphics[width=1\linewidth,trim=1.5cm 0.5cm 1.8cm 0cm,clip]{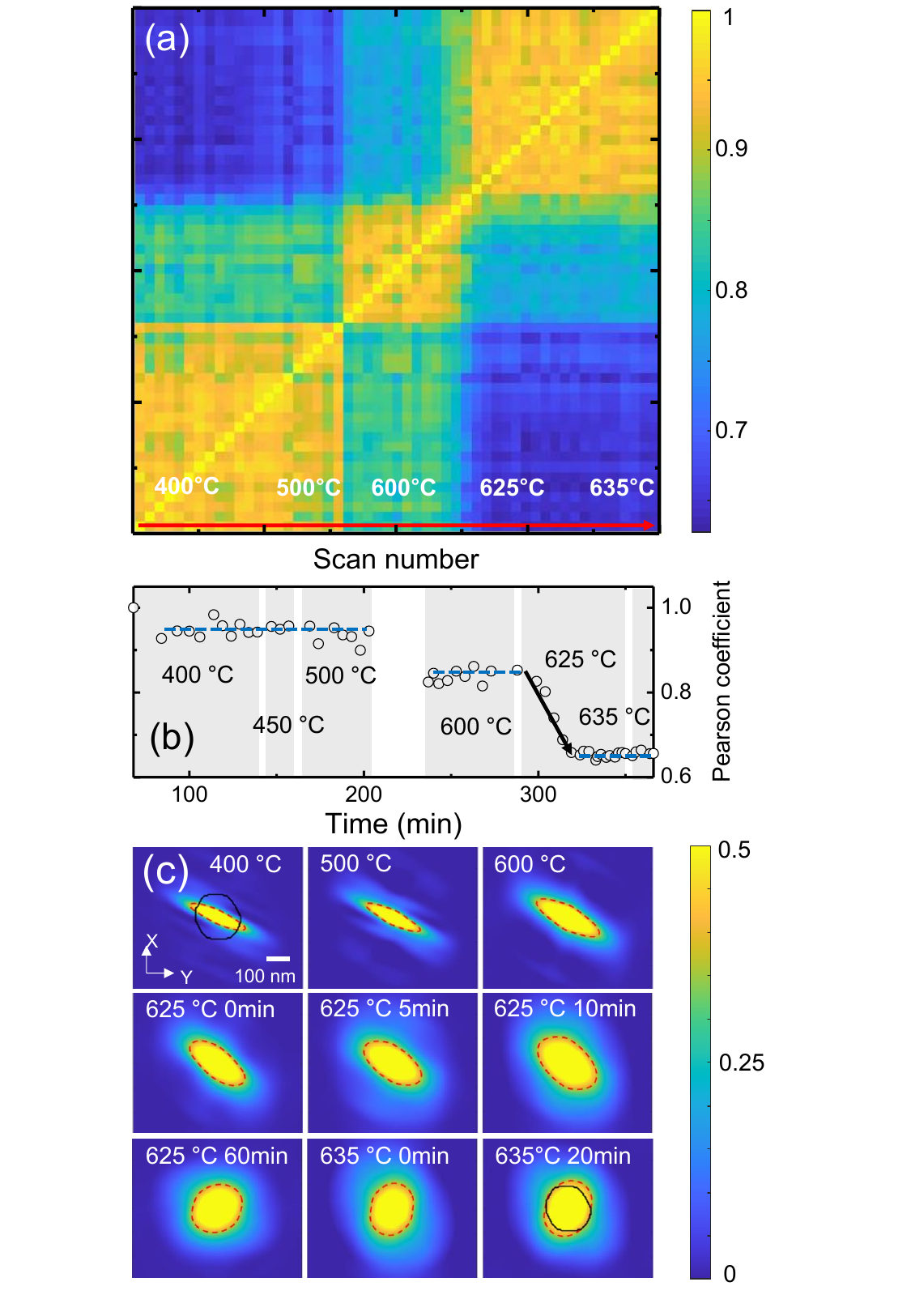}
    \caption{(a) Pearson correlation map derived from 3D diffraction datasets acquired during \textit{in-situ} annealing, illustrating the pairwise correlations across temperatures. The diagonal corresponds to the self-correlation of unity.
    (b) Line profile of the Pearson correlation coefficient along the red arrow line shown in (a). Significant structural changes occur at 600~$^\circ$C and 625~$^\circ$C. At 625~$^\circ$C, the correlation coefficient continuously changes with time.
    (c) \mbox{$Q_\mathrm{x}$--$Q_\mathrm{y}$} slice of the auto-correlation function (ACF) derived from 3D coherent diffraction data. The red dashed contour indicates the 1/\textit{e} drop of the ACF. The maximum of the color scale was set to 0.5. The overall domain morphology can be inferred from the ACF, which appears elongated initially and gradually thickens. At 625~$^\circ$C, pronounced temporal dynamics was observed as atoms gradually attach to the domain.
    }
    \label{fig:fig3}
\end{figure}

The changes in the widths of the diffraction profile along the $Q_z$ direction and the two directions in the \mbox{$Q_x$-$Q_y$} plane indicated in the second panel of Fig.~\ref{fig:fig2}(a), are shown in Fig.~\ref{fig:fig2}(c).
A noticeable decrease in the \mbox{$Q_{x\text{–}y}$} broad width observed above 400~$^\circ$C indicates that the crystalline domains gradually coarsened along the corresponding real-space direction, a process that became rapid above 600~$^\circ$C. 
The \mbox{$Q_z$} width also decreases, but only slightly, indicating that 
growth along the substrate normal is minimal.
In contrast, the $Q_{x\text{–}y}$ narrow width, which corresponds to the elongated direction in real space, remained largely unchanged at lower temperatures but increased slightly above 625~$^\circ$C.
This suggests that, beyond 625~$^\circ$C, where the faceting process became prominent, atomic rearrangement occurred as the initially hemispherical nanoparticle changed toward a faceted shape, resulting in a slight decrease in the real-space length along the initially elongated direction.

To examine the overall trend and identify the temperatures and time windows of structural evolution, we evaluated the Pearson correlation  coefficient between pairs of 3D diffraction datasets acquired sequentially in time and temperature, which is shown in Fig.~\ref{fig:fig3}(a).
The Pearson correlation coefficient is defined as:
\begin{equation} 
r(x,y)=  \frac{\sum_n \{(x_n - \bar{x}) \times (y_n-\bar{y})\}}{\sqrt{\sum_n(x_n-\bar{x})^2} \times \sqrt{\sum_n(y_n-\bar{y})^2}}~,
\end{equation}
where $x_n$ and $y_n$ represent individual elements in a pair of diffraction datasets being compared, and $\bar{x}$ and $\bar{y}$ denote the mean values of all elements in those respective datasets.
Upon exceeding the Tammann temperature ($\sim$400~$^\circ$C), atoms within a crystal domain are expected to acquire sufficient mobility to drive domain coarsening.
Between 400~$^\circ$C and 500~$^\circ$C, however, the diffraction pattern remains largely unchanged. 
At 600~$^\circ$C, the diffraction pattern changed instantaneously and remained steady, whereas at 625~$^\circ$C, it continued to evolve even after the temperature stabilization was complete.
This time-dependent dynamics observed at 625~$^\circ$C is indicated by the black arrow marking the reduction of the Pearson coefficient in Fig.~\ref{fig:fig3}(b) which shows the Pearson coefficient along the line marked in Fig.~\ref{fig:fig3}(a).
In this temperature region, the diffraction pattern continues to evolve with time, suggesting that domain coarsening driven by atomic rearrangement proceeds continuously.

Because the diffraction pattern continuously changes at 625~$^\circ$C even during measuring a single set of 3D diffraction profile, the measured pattern represents a temporal average. 
In such evolving regimes, standard iterative phase retrievals for BCDI fail to yield reliable reconstructions, making it difficult to capture the underlying structural dynamics.
To overcome this limitation, we analyzed the 3D autocorrelation function (ACF), which provides a reconstruction-free description of the correlation of the time-averaged real-space density.

Shown in Fig.~\ref{fig:fig3}(c) are the ACFs in the $Q_{x\text{--}y}$ plane obtained by direct Fourier transform of the measured diffraction intensity.
The dashed lines indicate the boundaries where the ACF decreases to $1/e$, providing an approximate shape of the major domain.
At 400~$^\circ$C, the ACF is elongated along a specific direction, with faint subsidiary peaks on each side of the central main peak. 
This indicates that, the main Au domain was initially elongated and there existed a nearby domain distinct from the main domain.
As the temperature increases to 600~$^\circ$C, the ACF gradually becomes more isotropic, and the subsidiary peaks are absorbed into the main peak.
At 625~$^\circ$C, the area of the main peak grows and changes continuously over time reflecting the growth of the main domain, which is consistent with the dynamic transition region observed in the Pearson correlation analysis.
Furthermore, by referring to the final-state morphology obtained by BCDI indicated by the black solid contour including the $\{111\}$ and $\{100\}$ facets (the BCDI reconstructed domain will be discussed later), 
we found that the Au domain grew primarily along the $[\mp2, \pm1, \pm1]$ directions at the expense of adjacent domains as the crystal approached towards its equilibrium shape.

\begin{figure}[t]
    \centering
    \includegraphics[width=1\linewidth,trim=0.2cm 5.5cm 0cm 0cm,clip]{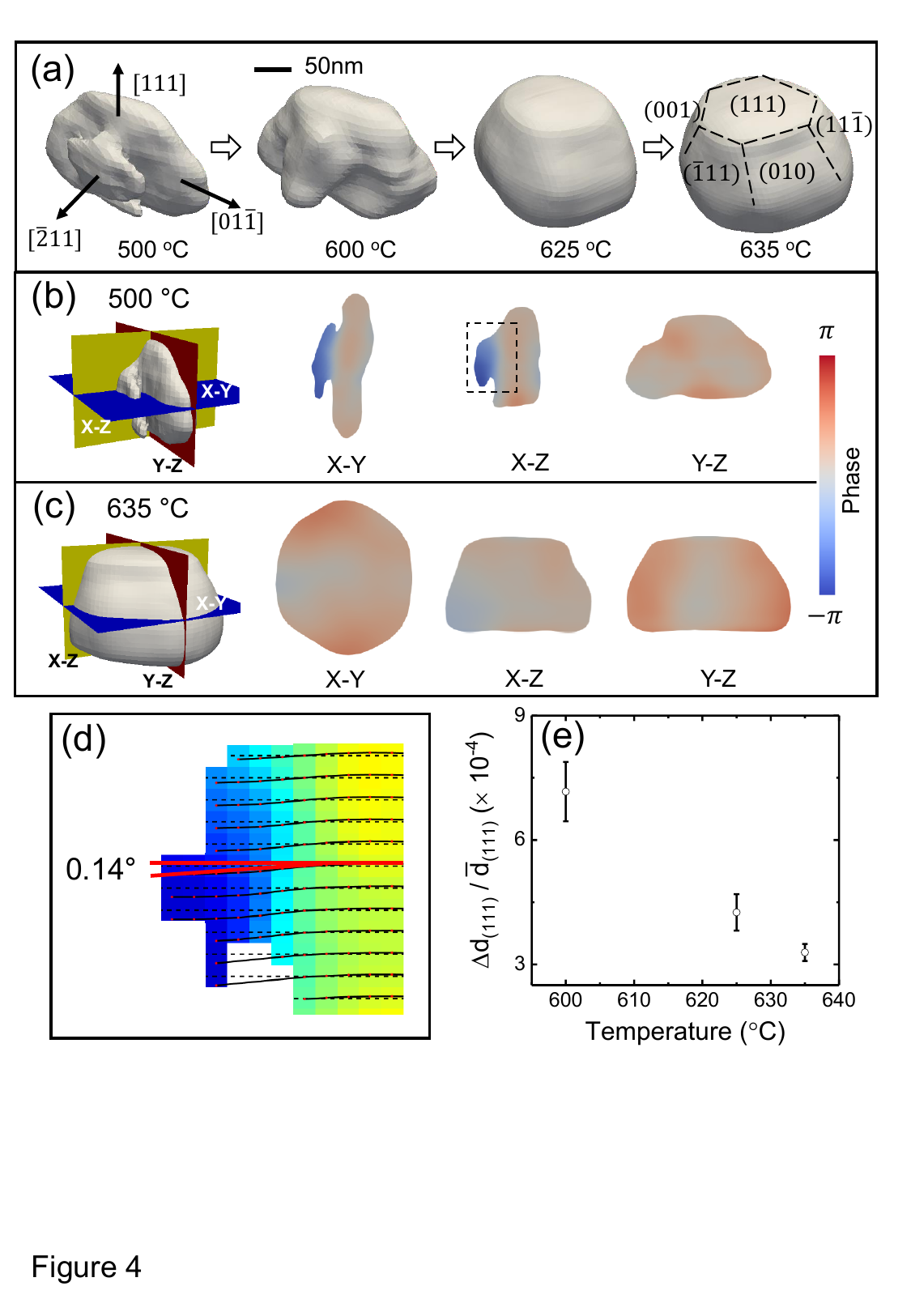}
    \caption{(a) Evolution of the domain morphology approaching a truncated octahedron at the final annealing at 635 $^\circ$C. 
    Cross-sectional phase maps obtained in the stabilized state at 500~$^\circ$C (b) and 635~$^\circ$C (c).
    A distinct phase gradient between the main and subsidiary domains is evident at 500~$^\circ$C, whereas
   the phase distribution becomes relatively uniform at 635~$^\circ$C, where crystalline facets have substantially developed.
    (d) Phase variation within the region indicated by a dashed box in (b). Phase values are mapped into the vertical voxel position, where a single voxel length corresponds to the $(111)$ d-spacing, $d_{(111)}$. The broken lines represent the reference (111) planes.
    (e) Distribution of $d_{(111)}$, extracted from the strain distribution within the reconstructed volume.}
    \label{fig:fig6}
\end{figure}

Finally, Fig.~\ref{fig:fig6}(a) displays 3D-rendered images of the domain reconstructed by BCDI in relatively stable regimes at a few annealing temperatures. 
The reconstructed images in Fig.~\ref{fig:fig6}(a) clearly exhibit the morphological evolution of the Au crystalline domain, which is consistent with the behavior of the XRD profiles and the ACF discussed above. 
During the initial low-temperature stage, the reconstructed Bragg electron density reveals an elongated primary domain, while surrounding domains diffracting outside the measured reciprocal space volume contribute negligibly.
The elongation is oriented approximately along the $[\pm1 10]$ directions, which correspond close-packed directions in FCC crystals.
This anisotropy may reflect the characteristics of the solidification during rapid cooling via laser quenching, in which the growth along the close-packed direction is preferred.
Since atoms are densely arranged in the $\left\{110\right\}$ planes, the kinetic growth during rapid cooling may favor elongation along the $<110>$ directions. 

As the temperature increases, the morphology gradually approaches the equilibrium truncated octahedral morphology.
At 600~$^\circ$C, the subsidiary domain merges with the main domain which becomes thicker progressively.
Upon increasing the temperature further
to 625~$^\circ$C, Au atoms continuously attach along the initially narrow direction of the domain, leading to the motion of the grain-boundary and the formation of a more isotropic volume. 
In the reconstruction result at 635~$^\circ$C, well-defined crystallographic facets characteristic of a truncated octahedron emerge; the domain shape is bounded by the $\left\{111\right\}$ and $\left\{100\right\}$ atomic plane families.

For a better understanding of the two-domain structure observed at 500~$^\circ$C, 
we analyzed the cross-sectional slices of the phase maps obtained from the BCDI reconstructions of the saturated diffraction data at 500~$^\circ$C and 635~$^\circ$C. 
As shown in the slices of Fig.~\ref{fig:fig6}(b), the phase of the main domain is relatively uniform, while the phase in the smaller subsidiary domain changes progressively as the outer boundary is approached.
The phase gradient in the region marked by a broken square in Fig.~\ref{fig:fig6}(b) likely originates from the tilt of the subsidiary domain \cite{mastropietro2017} relative to the main domain.
The variation of the phase values in this region is presented in Fig.~\ref{fig:fig6}(d). 
Here, the phase is mapped to the vertical position within each voxel, with the full length of a single voxel corresponds to the lattice displacement of one $(111)$ d-spacing.

For small crystal tilts, the tilt angle can be estimated from the phase gradient by ${\delta \phi}/{\delta x} = |\textbf{\textrm{G}}_{\textbf{111}}|\cdot \delta \textbf{u} / \delta x \simeq |\textbf{\textrm{G}}_{\textbf{111}}|\cdot \alpha$, where $\alpha$ is the angular separation of the (111) surface normals between the two crystals along the x direction (or [$\bar{2}$11] direction) and $\textbf{u}$ corresponds to the crystal displacement field.
The value of $\alpha$, evaluated from the phase map in this region, is about 0.14$^\circ$, which is comparable to the peak separation observed in the 400~$^\circ$C line profile in Fig.~\ref{fig:fig2}(b).
In contrast, the slices in Fig.~\ref{fig:fig6}(c) obtained at 635~$^\circ$C exhibit a relatively uniform phase distribution, similar to the main domain in Fig.~\ref{fig:fig6}(b).
This suggests that the atoms in the smaller neighboring domain migrated and redeposited onto the larger main domain, ultimately forming a single-crystalline nanoparticle---a process reminiscent of Ostwald ripening.
Annealing above 600 $^\circ$C reduces the variation in the (111) lattice spacing, as demonstrated by the temperature dependence of the strain distribution width in Fig.~\ref{fig:fig6}(e).
This implies that the crystalline order also increases as the morphology approaches an equilibrium truncated octahedral shape during the annealing process.

\begin{figure}[!t]
    \centering
    \includegraphics[width=1\linewidth,trim=0cm 16.5cm 0cm 0cm,clip]{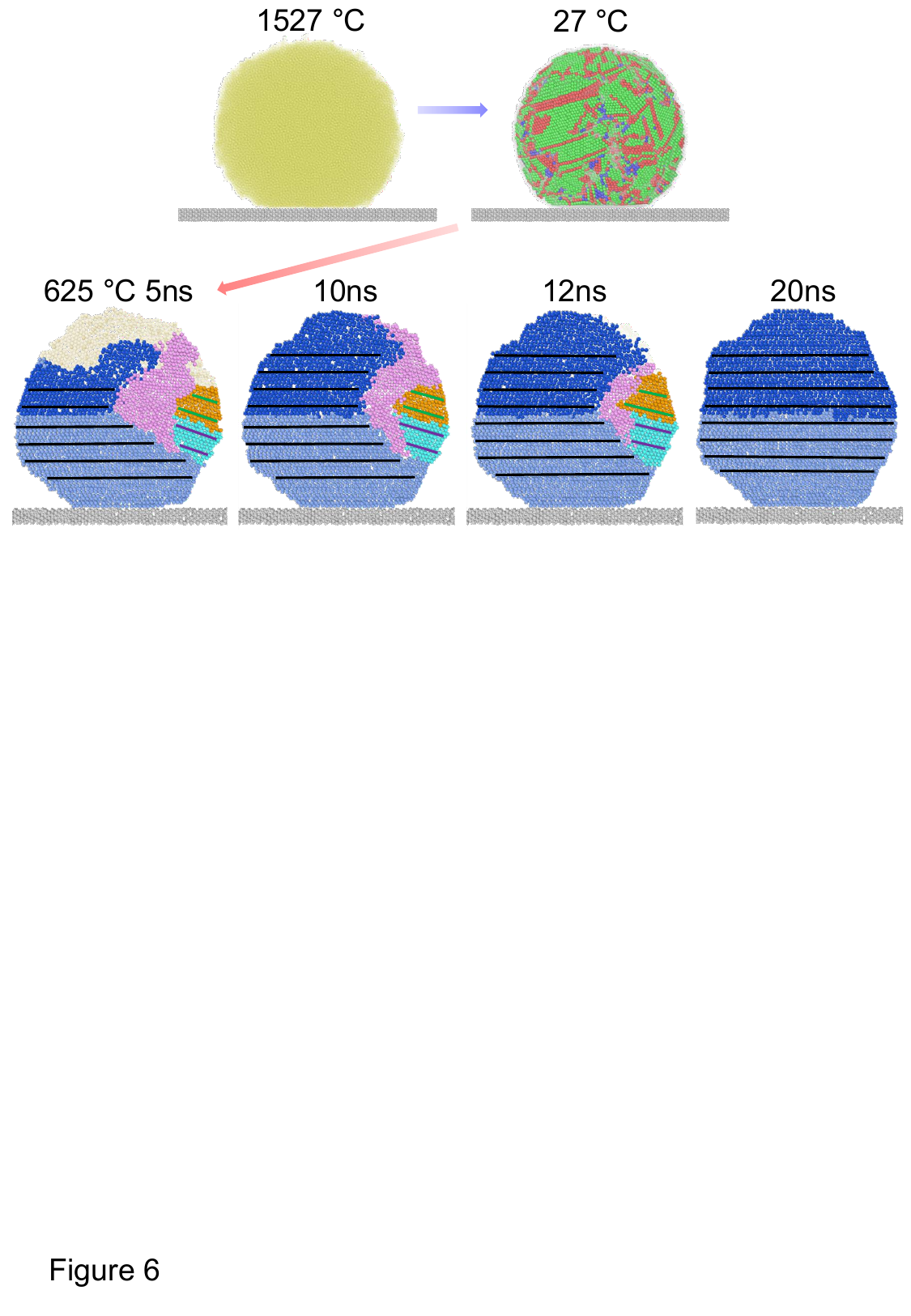}
    \caption{
    Results of MD simulations showing the crystalline domain evolution of $\sim$\,14\,nm AuNP.
    The upper panel shows the formation of a multi-domain structure (right) during rapid cooling from the liquid state (left), where the green, red, blue, and yellow represent regions with local FCC, HCP, BCC, and other structures, respectively.
    The lower panel presents cross-sectional views illustrating the domain evolution during annealing at 625~$^\circ$C. 
    The two large domains in blue and light-blue have their (111) planes aligned parallel to the substrate r-Al$_2$O$_3$ plane, indicated by the gray slab.
    Neighboring domains in yellow, pink and cyan have the (111) planes tilted away from the substrate plane. The black, green and purple lines indicate representative (111) planes within the selected domains.
    }
    
    \label{fig:fig7}
\end{figure}

The evolution of a $\sim$\,14\,nm multi-domain AuNP  was investigated via molecular dynamic (MD) simulations using LAMMPS employing an embedded-atom-method (EAM) potential \cite{ackland1987,thompson2022} (see the Supplemental Material for further details).
Despite the much smaller size and faster timescales are involved, the simulations exhibit the key features observed experimentally.
The laser quenching was simulated by dropping the temperature of the AuNP in liquid state at 1527~$^\circ$C to 27~$^\circ$C instantaneously, which resulted a multi-domain structure composed of small domains with diverse structures and orientations as shown in the upper panel of Fig.~\ref{fig:fig7}.

The lower panel of Fig.~\ref{fig:fig7} shows the evolution of the Au particle during annealing at 625~$^\circ$C.
Upon annealing for 5 ns, a few selected domains coarsen while smaller domains disappear.
The large blue and light blue domains both have their (111) planes nearly parallel to the substrate, but possess distinct in-plane orientations.
Neighboring yellow, pink, and cyan domains are distinguished by the orientations of their (111) plane which are tilted away from the substrate normal.
Up to 12\,ns, the largest two domains grow across the domain boundary at the expense of the neighboring domains indicating grain-boundary migration.
Atoms near the boundaries rearrange locally, and the boundary migrates toward the smaller domain.
With further annealing to 20\,ns, the large two domains continue to grow occupying a substantial portion of the particle boundary, while their crytstallinity improve significantly.

The simulations shows that after the initial coarsening of large grains via grain boundary migration, the domain structure evolved more gradually.
In fact, even after 20 ns, a few domains persist suggesting the coarsening of larger domains requires substantially more time.
The slower evolution may be associated with energy barriers to the migration of atoms from relatively large remaining domains.
The domain growth observed experimentally over several minutes during the early stages at 625~$^\circ$C may reflect slower coarsening of the remaining larger domains.

In this letter, we investigated the structural evolution of nanocrystalline domains in a laser-quenched AuNP during \emph{in-situ} annealing using 3D coherent X-ray diffraction.
BCDI and ACF analyses revealed that initially, two crystalline domains contribute to the Au(111) reflection, which gradually merge and coarsen into a single-domain crystal during annealing.
Pearson correlation analysis identified three stages of evolution: domain merging above 600~$^\circ$C, continuous coarsening toward a more isotropic morphology at 625~$^\circ$C, and the emergence of low-energy $\left\{111\right\}$ and $\left\{100\right\}$ facets approaching a truncated octahedral shape at 635~$^\circ$C.
MD simulations provided complementary quantitative atomistic insights into the domain-growth process.
This work demonstrates the potential of combining ACF analysis and BCDI reconstruction to resolve the three-dimensional evolution of nanocrystalline domains under dynamic conditions, providing insight into the mechanisms of domain growth and equilibrium shape formation.

\textit{Acknowledgements}---We thank Jung Won Choi for sample and laser preparation, Ross Harder for experimental support, and Juncheng E for guidance on the MD simulations.
This work was supported by the National
Research Foundation of Korea (NRF) by grants RS-2022-00154676, RS-2025-02318077, and RS-2026-25491663.
This research used resources of the Advanced Photon Source, a U.S. Department of Energy (DOE) Office of Science User Facility operated by Argonne National Laboratory under Contract No. DE-AC02-06CH11357.

\newpage

\section{Supporting Information}

\setcounter{figure}{0}
\setcounter{table}{0}
\setcounter{equation}{0}
\renewcommand{\thefigure}{S\arabic{figure}}
\renewcommand{\thetable}{S\arabic{table}}
\renewcommand{\theequation}{S\arabic{equation}}
\renewcommand{\thesection}{S\arabic{section}}
\renewcommand{\thesubsection}{\thesection.\arabic{subsection}}

\subsection{Electron microscopy characterization}
\begin{figure}[!htbp]
    \centering    \includegraphics[width=1\linewidth,trim=0cm 17.3cm 0cm 0.5cm,clip]{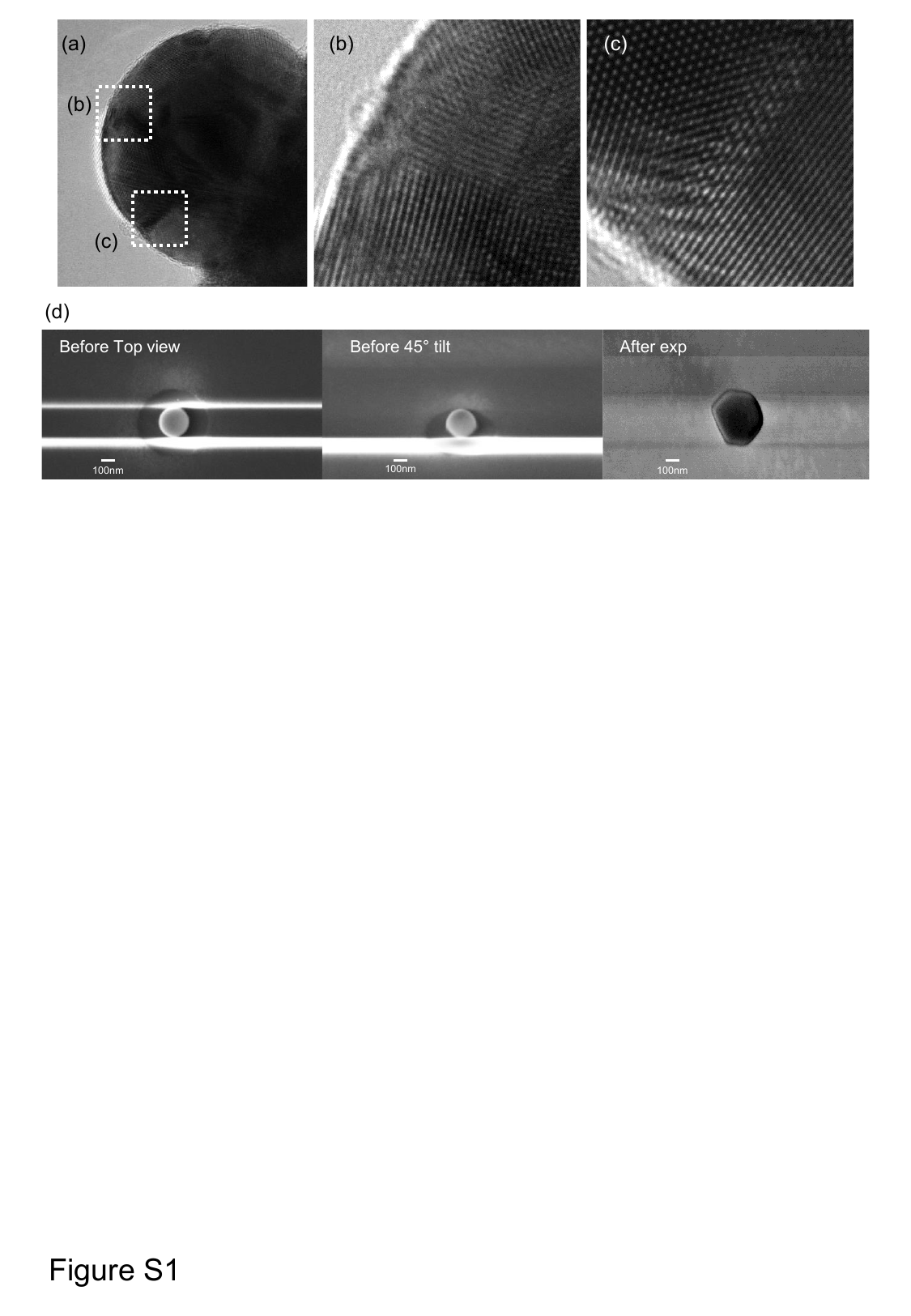}
    \caption{Electron microscopy images of the fabricated Au nanoparticles. (a) TEM image of a single nanoparticle. (b,c) High-resolution TEM images taken from the regions marked in (a), showing multidomain structure.
    (d) Top-view SEM image (left), 45$^\circ$ tilted SEM image (middle), confirming the spherical outer morphology of the particle, and top-view SEM image of the final product after \textit{in-situ} experiment (right).
    }
    \label{fig:figs1}
\end{figure}

Isolated, multi-domain AuNPs were designed and fabricated using the following procedure.
First, a 15\,nm thick Au thin film was deposited by e-beam evaporation on a r-plane Al$_2$O$_3$ $(1\bar{1}02)$ substrate masked by a 2000 mesh grid.
The film was annealed at 700~$^\circ$C for 3 hours to obtain NPs via solid-phase dewetting process. 
Small debris particles and residuals were removed by Ar ion milling, followed by further annealing at 600~$^\circ$C for 4 hours.
The AuNPs resulted from the above solid-phase dewetting are typically single-domain crystals with well-defined crystal shape \cite{hwang2013}.

Thus prepared AuNP crystals were irradiated with an intense single laser pulse at a wavelength of 532\,nm, a pulse duration of $\sim$2\,ns, and a fluence of 0.17\,J/cm².
The laser irradiated AuNPs undergo partial melting and rapidly re-solidify into polycrystalline, hemispherical NPs.
This rapid cooling typically creates internal defects, such as twin boundaries and grain boundaries, forming a multidomain structure with various orientations coexisting within a single particle \cite{huang2023,lasemi2024,link2003,huang2021,kim2020}.

As revealed by the SEM and TEM images in Fig.~\,S1, the AuNPs investigated in this study exhibit a rounded, nearly hemispherical morphology, yet they are polycrystalline, comprising multiple crystal domains oriented in different directions.
A nanoparticle with a rounded morphology and an internal multidomain structure was selected as the starting state. We then traced the evolution from the multidomain to a single-domain structure during \textit{in-situ} annealing, along with the accompanying change of the domain morphology.

\newpage
\subsection{BCDI setup schematic}
\begin{figure}[!htbp]
    \centering    \includegraphics[width=1\linewidth,trim=0cm 14cm 0cm 0cm,clip]{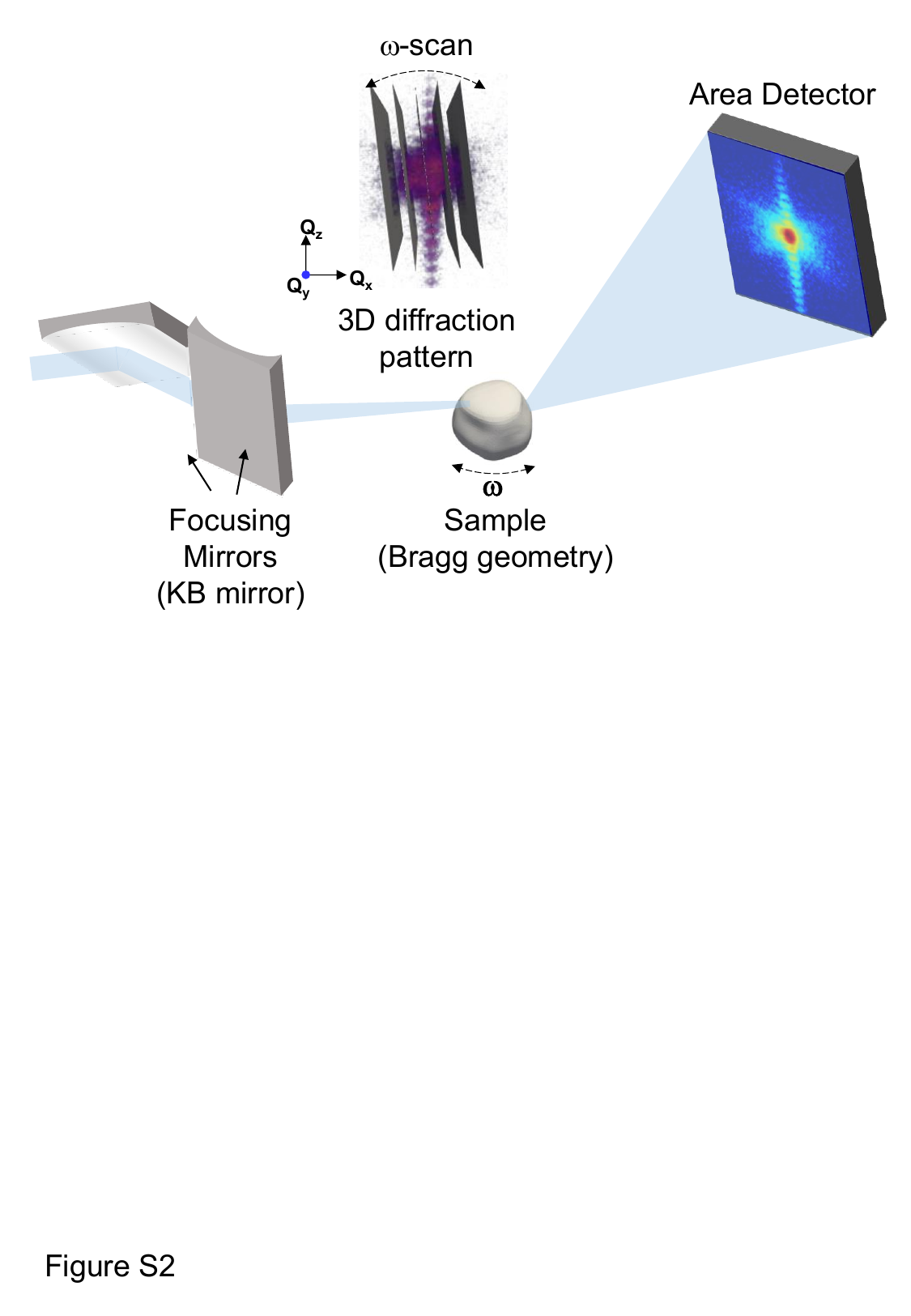}
    \caption{Schematic of the Bragg coherent diffraction imaging setup. A coherent X-ray beam, focused by a Kirkpatrick–Baez (KB) mirror, illuminates an Au nanocrystal satisfying the Bragg condition. The diffraction signal is collected during an $\omega$-scan to enable 3D image reconstruction of the nanocrystal.}
    \label{fig:figs2}
\end{figure}

The BCDI measurement was conducted at the \mbox{34-ID-C} beamline of the Advanced Photon Source, Argonne National Laboratory, USA.
Coherent X-ray beam at 10\,keV, focused with a Kirkpatrick-Baez mirror to $\sim$\,700~nm\,$\times$\,700~nm, was used to obtain diffraction patterns around the Au(111) Bragg reflection from an isolated nano-scale Au particle.
Diffraction patterns were recorded using a Timepix detector with 55\,$\mu m$\,$\times$\,55\,$\mu m$ sized pixels, positioned 0.5\,m downstream of the sample. At each temperature and time point, a total of 61 frames were acquired during the rocking of the specimen with a step size of 0.01$^\circ$.
A schematic illustration of the diffraction measurement setup is provided in Fig.~\,S2.
The sample temperature was varied from room temperature to 635~$^\circ$C stepwise inside a heating chamber under a nitrogen atmosphere. 
Because the diffraction patterns exhibited no significant temporal change over the duration required to collect a full dataset, we adopted a stepwise approach to increase the temperature to 400~$^\circ$C.
Above this threshold, we continuously monitored and analyzed the time-dependent changes in the diffraction patterns to observe their evolution at selected temperatures.
At each temperature, the 3D coherent diffraction patterns in the vicinity of the (111) Bragg peak were repeatedly recorded until no further significant changes were observed.

The phase retrieval procedure comprised of a total of 2220 iterations using a combination of the hybrid input--output (HIO) and the error reduction (ER) algorithms \cite{fienup1982,robinson2009,maddali2019}. 
Specifically, the ER algorithm was first applied for 20 iterations, followed by 180 iterations of the HIO algorithm. 
This ER--HIO combination was repeated for 11 rounds and the reconstruction was concluded with an additional 20 ER iterations. 
During the iterative reconstruction process, a partial coherence correction was incorporated \cite{clark2012}. 
The real space voxel size of the reconstructed images was approximately 7.8\,nm. 
The 3D reconstructed Bragg amplitude and phase maps were visualized using the Paraview software (http://www.paraview.org).
The isosurface threshold was set to 0.3 by comparing the nanoparticle size observed in SEM images with that of the reconstructed domain after the annealing was completed at a final temperature of 635~$^\circ$C (Fig.~S1(d)).

\subsection{Additional ACF slices}
\begin{figure}[!htbp]
    \centering    \includegraphics[width=1\linewidth,trim=0cm 0.7cm 0cm 0cm,clip]{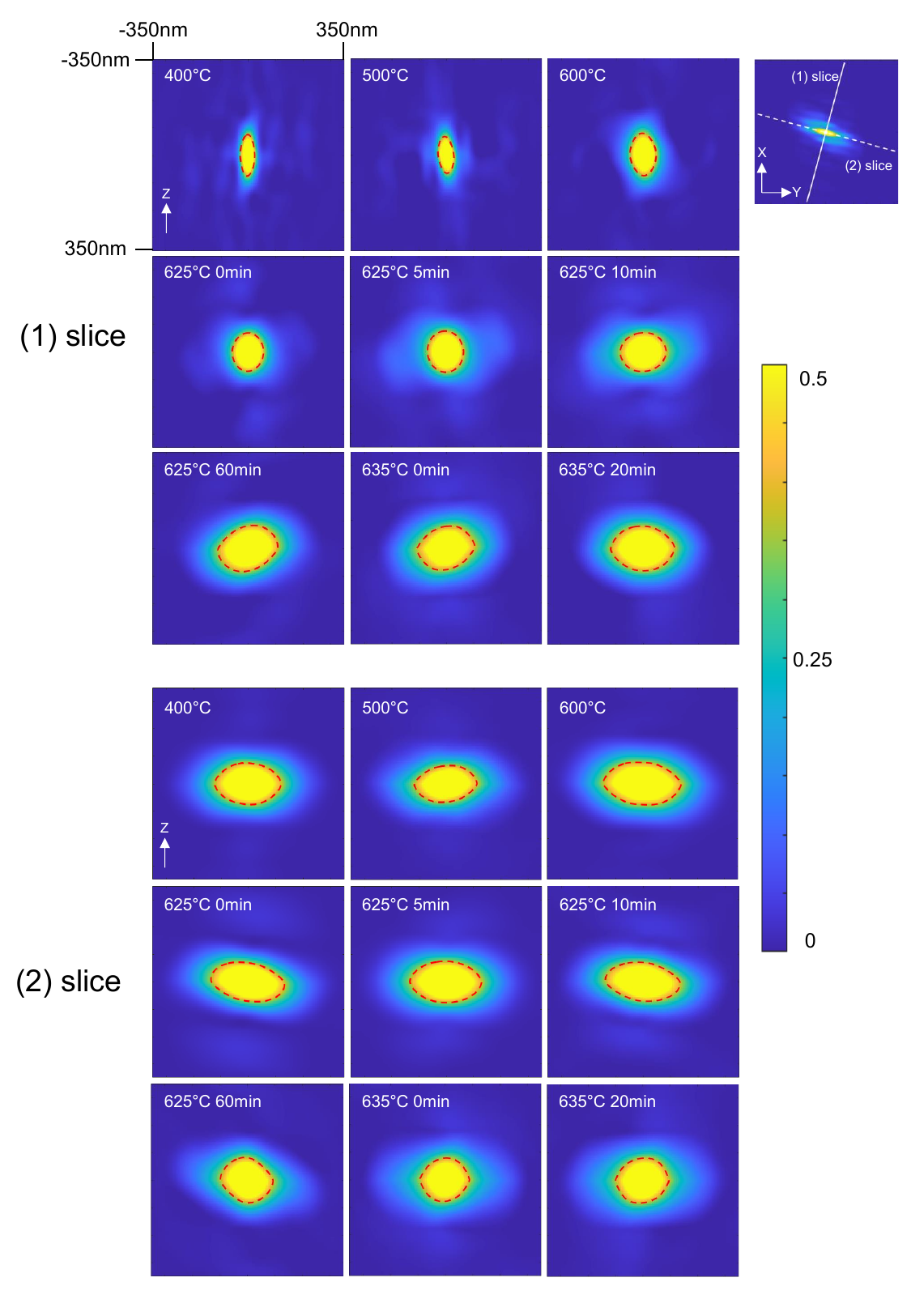}
    \caption{Additional ACF slice maps including the $Q_z$ direction. The slice orientations are indicated by the inset on the top right.}
    \label{fig:figs3}
\end{figure}

\newpage
\subsection{ACF volume graph}
\begin{figure}[!htbp]
    \centering    \includegraphics[width=1\linewidth,trim=0.5cm 13.5cm 0.5cm 1cm,clip]{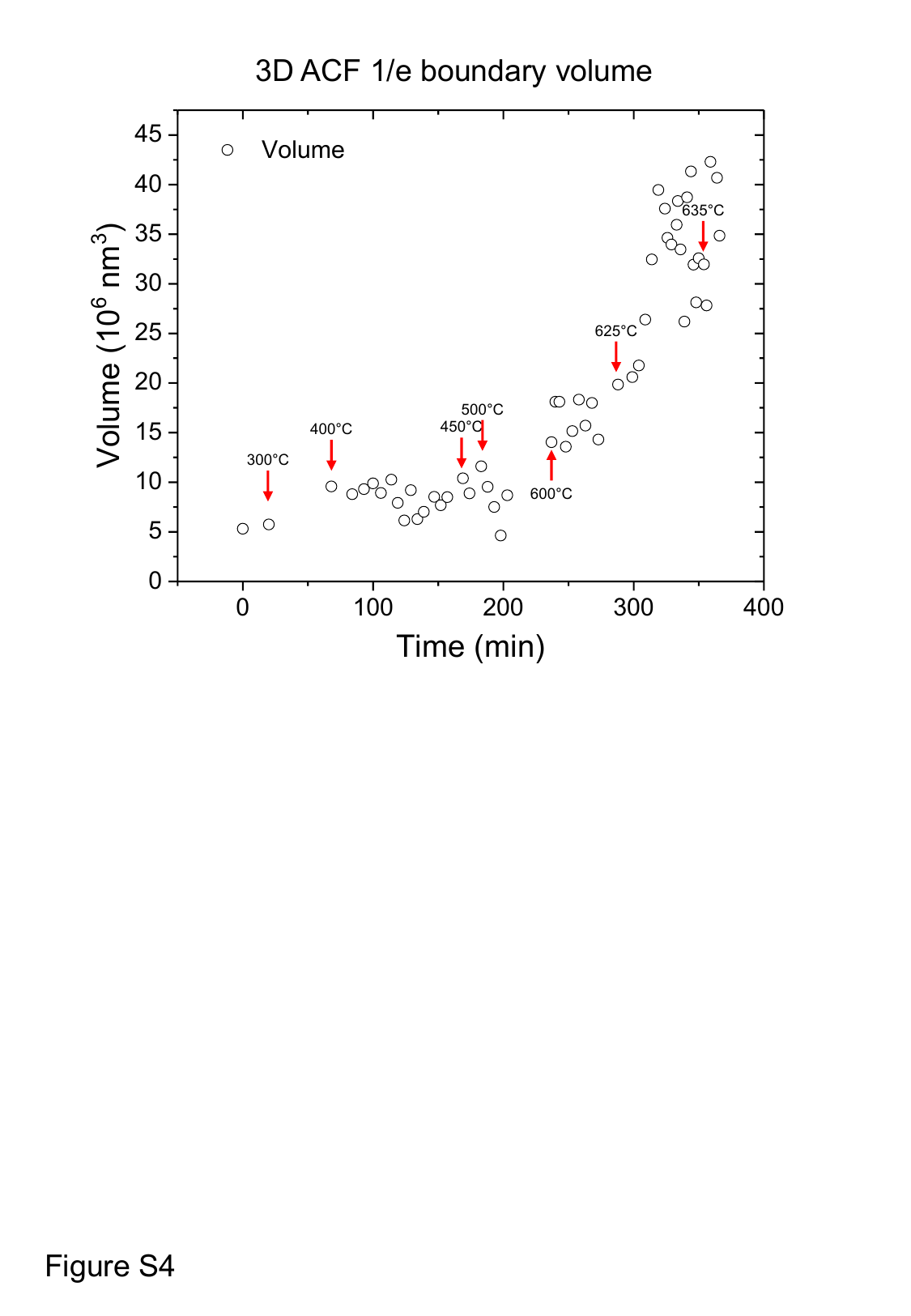}
    \caption{Volume evolution as a function of annealing temperature and time. 3D ACF volume calculated from the 1/e boundary during the total experiment procedure. The red arrows indicate the point in time when the temperature changed.}
    \label{fig:figs4}
\end{figure}

\subsection{3D rendering image at 635 $^\circ$C}
\begin{figure}[!htbp]
    \centering    \includegraphics[width=1\linewidth,trim=0cm 21cm 0.5cm 0cm,clip]{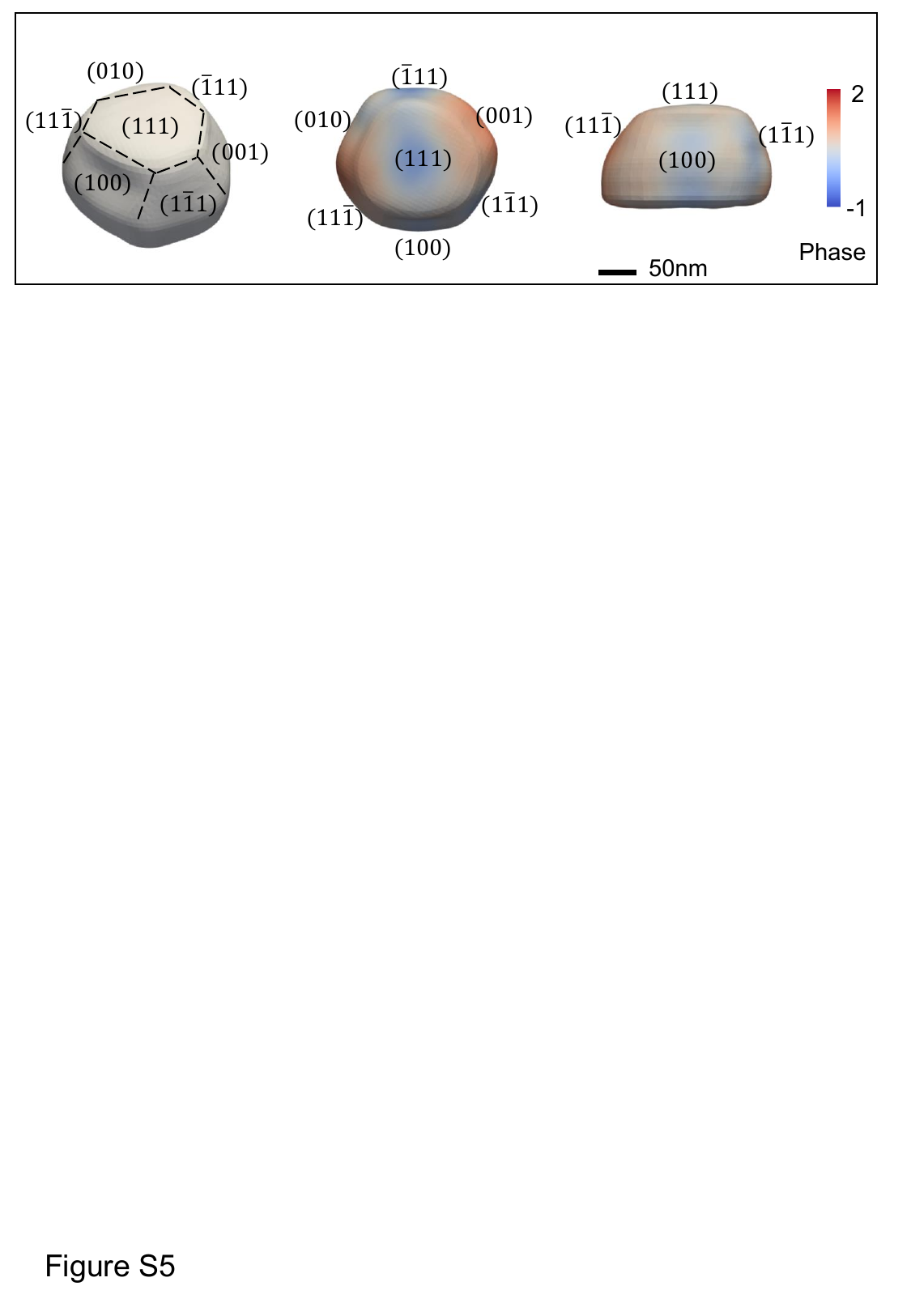}
    \caption{Domain morphology resembling a truncated octahedron, observed at the final annealing stage of 635 $^\circ$C. Crystalline facets are labeled, and phase values are displayed in both the top and side views of the domain.
    }
    \label{fig:figs5}
\end{figure}

The morphology of the AuNP domain in the final stage is illustrated in Fig.~\,S5, which clearly shows the crystal facets.
The indices of the facets are identified based on the angular relationships between the facet normals and the [111] direction.
In an fcc crystal, the angles between the $(111)$ and $\left\{100\right\}$ planes and between the $(111)$ and $\left\{11\bar{1}\right\}$ planes are 54.7$^\circ$  and 70.5$^\circ$, respectively.
As the low-energy $\left\{111\right\}$ and $\left\{100\right\}$ facets expanded, the crystalline domain morphology approaches  the equilibrium octahedron shape.

\clearpage
\subsection{MD simulation results}

MD simulations were performed using LAMMPS with the Ackland EAM potential for Au. 
The nanoparticle consisted 94,874 atoms with a size of approximately 14\,nm. An r-plane Al$_2$O$_3$ substrate was placed beneath the nanoparticle which parallel to the $xy$ plane. The Al and O atoms were fixed. The Au--Al and Au--O interactions were modeled using a Lennard--Jones pair potential with parameters $\epsilon = 0.0067$\,eV and $\sigma = 3.2$\,\AA, with a cutoff distance of $10.0$\,\AA.
The temperature was controlled using a Nos\'e--Hoover thermostat.
The time step for integrating the equations of motion was 1\,fs during the initial stages and was increased to 3\,fs during the later annealing stages after numerical stability checks to reduce the computation time.

The initial crystalline truncated octahedron nanoparticle was annealed to 1800\,K (1527~$^\circ$C) and dwell for 100\,ps. The atomic volume--temperature hysteresis curves obtained during heating and cooling supported the formation of a liquid state at this temperature. 
The nanoparticle was then cooled to 300\,K (27~$^\circ$C) at a rate of 1.6\,K/ps.
The resulting multi-domain nanoparticle was annealed to 898\,K (625~$^\circ$C) at a rate of 0.4\,K/ps and dwell at this temperature for 20\,ns.

\newpage
\begin{figure}[!t]
    \centering    \includegraphics[width=0.96\linewidth,trim=0cm 1.5cm 0cm 0cm,clip]{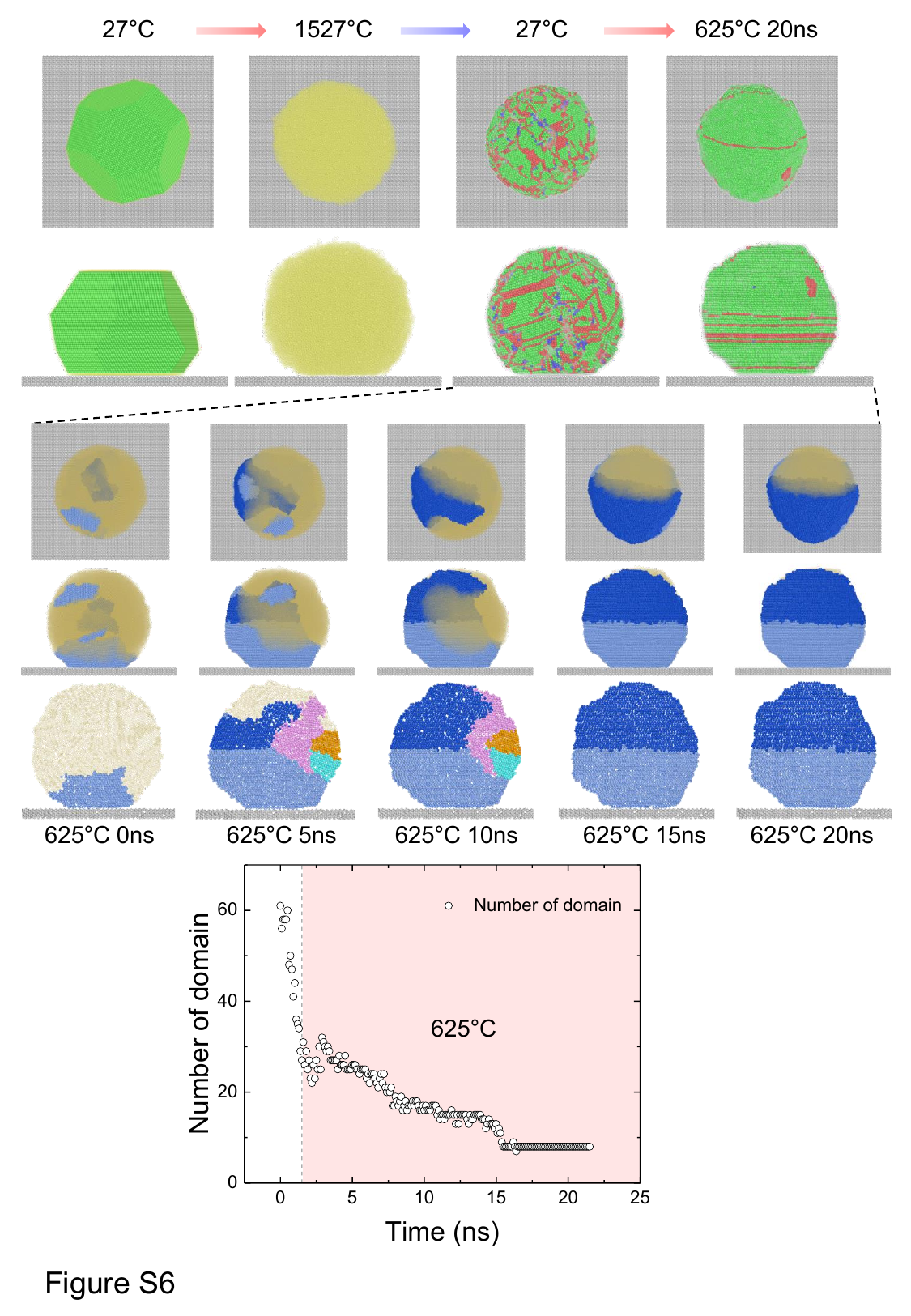}
    \caption{Overall MD simulations of multi--domain formation and coarsening in an Au nanoparticle.
    The upper two rows show top and side views of the well defined nanocrystal at 27~$^\circ$C, melting at 1527~$^\circ$C, rapid cooling to 27~$^\circ$C, and subsequent annealing at 625~$^\circ$C for 20\,ns.
    Atoms are colored according to the local structures identified using PTM with green, red, blue, and yellow which indicate FCC, HCP, BCC, and others, respectively. 
    The following two rows show the corresponding top and side views of domain evolution during annealing at 625~$^\circ$C from 0\,ns to 20\,ns.
    The largest domain in the final state at 625~$^\circ$C after 20\,ns is traced back and colored with blue.
    Other domains with similar (111) plane normal is nearly aligned with substrate normals but different in-plane orientations are show in light blue.
    The remaining Au atoms are colored with transparent yellow.
    The lowest row shows atomic-scale cross sections. The main domain highlighted in blue, while other domain whose (111) plane normals are nearly aligned with the substrate normal are shown in light blue, same in the rows above. Neighboring domains with differently oriented (111) plane normals are shown in pink, orange, and cyan.
    The horizontal gray atoms represent the r-plane $Al_2O_3$ substrate.
    The bottom graph shows the number of domains using grain segmentation as a function of annealing time.}
    \label{fig:figs6}
\end{figure}

\newpage

\bibliographystyle{apsrev4-2}
\bibliography{references}

@article{carone2023,
  title = {Gold Nanoparticle Shape Dependence of Colloidal Stability Domains},
  author = {Carone, Antonio and Emilsson, Samuel and Mariani, Pablo and D{\'e}sert, Anthony and Parola, Stephane},
  year = 2023,
  journal = {Nanoscale Advances},
  volume = {5},
  number = {7},
  pages = {2017--2026},
  doi = {10.1039/D2NA00809B}
}

@article{daniel2004,
  title = {Gold {{Nanoparticles}}: {{Assembly}}, {{Supramolecular Chemistry}}, {{Quantum-Size-Related Properties}}, and {{Applications}} toward {{Biology}}, {{Catalysis}}, and {{Nanotechnology}}},
  shorttitle = {Gold {{Nanoparticles}}},
  author = {Daniel, Marie-Christine and Astruc, Didier},
  year = 2004,
  month = jan,
  journal = {Chemical Reviews},
  volume = {104},
  number = {1},
  pages = {293--346},
  doi = {10.1021/cr030698+}
}

@article{dykman2012,
  title = {Gold Nanoparticles in Biomedical Applications: Recent Advances and Perspectives},
  shorttitle = {Gold Nanoparticles in Biomedical Applications},
  author = {Dykman, Lev and Khlebtsov, Nikolai},
  year = 2012,
  journal = {Chem. Soc. Rev.},
  volume = {41},
  number = {6},
  pages = {2256--2282},
  doi = {10.1039/C1CS15166E}
}

@article{dzhigaev2022,
  title = {Three-Dimensional in Situ Imaging of Single-Grain Growth in Polycrystalline {{In2O3}}:{{Zr}} Films},
  shorttitle = {Three-Dimensional in Situ Imaging of Single-Grain Growth in Polycrystalline {{In2O3}}},
  author = {Dzhigaev, Dmitry and Smirnov, Yury and Repecaud, Pierre-Alexis and Mar{\c c}al, Lucas Atila Bernardes and Fevola, Giovanni and Sheyfer, Dina and Jeangros, Quentin and Cha, Wonsuk and Harder, Ross and Mikkelsen, Anders and Wallentin, Jesper and {Morales-Masis}, Monica and Stuckelberger, Michael Elias},
  year = 2022,
  month = jun,
  journal = {Communications Materials},
  volume = {3},
  number = {1},
  pages = {38},
  doi = {10.1038/s43246-022-00260-4}
}

@article{elbaum1993,
  title = {Relation of Growth and Equilibrium Crystal Shapes},
  author = {Elbaum, M. and Wettlaufer, J. S.},
  year = 1993,
  month = oct,
  journal = {Physical Review E},
  volume = {48},
  number = {4},
  pages = {3180--3183},
  doi = {10.1103/PhysRevE.48.3180}
}

@article{fienup1982,
  title = {Phase Retrieval Algorithms: A Comparison},
  shorttitle = {Phase Retrieval Algorithms},
  author = {Fienup, J. R.},
  year = 1982,
  month = aug,
  journal = {Applied Optics},
  volume = {21},
  number = {15},
  pages = {2758},
  doi = {10.1364/AO.21.002758}
}

@article{haruta1997,
  title = {Size- and Support-Dependency in the Catalysis of Gold},
  author = {Haruta, Masatake},
  year = 1997,
  month = apr,
  journal = {Catalysis Today},
  volume = {36},
  number = {1},
  pages = {153--166},
  doi = {10.1016/S0920-5861(96)00208-8}
}

@article{hu2025,
  title = {Optimization of Single Crystal Surface and Interface Structures for Electrocatalysis},
  author = {Hu, Haixiao and Liang, Haiyan and Liu, Xiaoyan and Jiang, Hehe and Yi, Moyu and Wu, Yongzhong and Hao, Xiaopeng and Chang, Bin and Zhou, Weijia},
  year = 2025,
  month = aug,
  journal = {Materials Reports: Energy},
  volume = {5},
  number = {3},
  pages = {100358},
  doi = {10.1016/j.matre.2025.100358}
}

@article{huang2021,
  title = {Atomistic {{View}} of {{Laser Fragmentation}} of {{Gold Nanoparticles}} in a {{Liquid Environment}}},
  author = {Huang, Hao and Zhigilei, Leonid V.},
  year = 2021,
  month = jun,
  journal = {The Journal of Physical Chemistry C},
  volume = {125},
  number = {24},
  pages = {13413--13432},
  doi = {10.1021/acs.jpcc.1c03146}
}

@article{huang2023,
  title = {Formation of {{Twin Boundaries}} in {{Rapidly Solidified Metals}} through {{Deformation Twinning}}},
  author = {Huang, Binting and Yang, Jishi and Luo, Zhiheng and Wang, Yang and Wang, Nan},
  year = 2023,
  month = jun,
  journal = {Materials},
  volume = {16},
  number = {13},
  pages = {4503},
  doi = {10.3390/ma16134503}
}

@article{hutchings2005,
  title = {Catalysis by Gold},
  author = {Hutchings, Graham J.},
  year = 2005,
  month = feb,
  journal = {Catalysis Today},
  volume = {100},
  number = {1-2},
  pages = {55--61},
  doi = {10.1016/j.cattod.2004.12.016}
}

@article{ishida2020,
  title = {Importance of {{Size}} and {{Contact Structure}} of {{Gold Nanoparticles}} for the {{Genesis}} of {{Unique Catalytic Processes}}},
  author = {Ishida, Tamao and Murayama, Toru and Taketoshi, Ayako and Haruta, Masatake},
  year = 2020,
  month = jan,
  journal = {Chemical Reviews},
  volume = {120},
  number = {2},
  pages = {464--525},
  doi = {10.1021/acs.chemrev.9b00551}
}

@article{bachmann2025,
  title={Coherent X-ray Diffraction Imaging of a Twinned PtRh Catalyst Nanoparticle under Operando Conditions},
  author={Bachmann, Lydia J and Lapkin, Dmitry and Schober, Jan-Christian and Dolling, Daniel Silvan and Kim, Young Yong and Assalauova, Dameli and Mukharamova, Nastasia and Dwivedi, Jagrati and Schulli, Tobias U and Keller, Thomas F and others},
  journal={ACS nano},
  year={2025},
  volume = {19},
  pages = {23552-23563},
  publisher={ACS Publications}
}

@article{kim2018,
  title = {Three-{{Dimensional Imaging}} of {{Phase Ordering}} in an {{Fe-Al Alloy}} by {{Bragg Ptychography}}},
  author = {Kim, Chan and Chamard, Virginie and Hallmann, J{\"o}rg and Roth, Thomas and Lu, Wei and Boesenberg, Ulrike and Zozulya, Alexey and Leake, Steven and Madsen, Anders},
  year = 2018,
  month = dec,
  journal = {Physical Review Letters},
  volume = {121},
  number = {25},
  pages = {256101},
  doi = {10.1103/PhysRevLett.121.256101}
}

@article{yau2017,
  title={Bragg coherent diffractive imaging of single-grain defect dynamics in polycrystalline films},
  author={Yau, Allison and Cha, Wonsuk and Kanan, Matthew W and Stephenson, G Brian and Ulvestad, Andrew},
  journal={Science},
  volume={356},
  number={6339},
  pages={739--742},
  year={2017},
  publisher={American Association for the Advancement of Science}
}

@article{lasemi2024,
  title = {Defect-{{Rich CuZn Nanoparticles}} for {{Model Catalysis Produced}} by {{Femtosecond Laser Ablation}}},
  author = {Lasemi, Niusha and Wicht, Thomas and Bernardi, Johannes and Liedl, Gerhard and Rupprechter, G{\"u}nther},
  year = 2024,
  month = jul,
  journal = {ACS Applied Materials \& Interfaces},
  volume = {16},
  number = {29},
  pages = {38163--38176},
  doi = {10.1021/acsami.4c07766}
}

@article{liang2022,
  title = {Size-{{Dependent Catalytic Behavior}} of {{Gold Nanoparticles}}},
  author = {Liang, Chen and Cheong, Jun Young and Sitaru, Gabriel and Rosenfeldt, Sabine and Schenk, Anna S. and Gekle, Stephan and Kim, II-Doo and Greiner, Andreas},
  year = 2022,
  month = feb,
  journal = {Advanced Materials Interfaces},
  volume = {9},
  number = {4},
  pages = {2100867},
  doi = {10.1002/admi.202100867}
}

@article{lim2022,
  title = {Unraveling the {{Simultaneous Enhancement}} of {{Selectivity}} and {{Durability}} on {{Single}}-{{Crystalline Gold Particles}} for {{Electrochemical CO}}{\textsubscript{2}} {{Reduction}}},
  author = {Lim, Yun Ji and Seo, Dongho and Abbas, Syed Asad and Jung, Haeun and Ma, Ahyeon and Lee, Kug-Seung and Lee, Gaehang and Lee, Hosik and Nam, Ki Min},
  year = 2022,
  month = jul,
  journal = {Advanced Science},
  volume = {9},
  number = {20},
  pages = {2201491},
  doi = {10.1002/advs.202201491}
}

@article{link2003,
  title = {Optical {{Properties}} and {{Ultrafast Dynamics}} of {{Metallic Nanocrystals}}},
  author = {Link, Stephan and {El-Sayed}, Mostafa A.},
  year = 2003,
  month = oct,
  journal = {Annual Review of Physical Chemistry},
  volume = {54},
  number = {1},
  pages = {331--366},
  doi = {10.1146/annurev.physchem.54.011002.103759}
}

@article{maddali2019,
  title = {Phase Retrieval for {{Bragg}} Coherent Diffraction Imaging at High X-Ray Energies},
  author = {Maddali, S. and Allain, M. and Cha, W. and Harder, R. and Park, J.-S. and Kenesei, P. and Almer, J. and Nashed, Y. and Hruszkewycz, S. O.},
  year = 2019,
  month = may,
  journal = {Physical Review A},
  volume = {99},
  number = {5},
  pages = {053838},
  doi = {10.1103/PhysRevA.99.053838}
}

@article{pacchioni2018,
  title = {Controlling the Charge State of Supported Nanoparticles in Catalysis: Lessons from Model Systems},
  shorttitle = {Controlling the Charge State of Supported Nanoparticles in Catalysis},
  author = {Pacchioni, Gianfranco and Freund, Hans-Joachim},
  year = 2018,
  journal = {Chemical Society Reviews},
  volume = {47},
  number = {22},
  pages = {8474--8502},
  doi = {10.1039/C8CS00152A}
}

@article{park2024,
  title = {In-Situ and Wavelength-Dependent Photocatalytic Strain Evolution of a Single {{Au}} Nanoparticle on a {{TiO2}} Film},
  author = {Park, Sung Hyun and Kim, Sukyoung and Park, Jae Whan and Kim, Seunghee and Cha, Wonsuk and Lee, Joonseok},
  year = 2024,
  month = jun,
  journal = {Nature Communications},
  volume = {15},
  number = {1},
  pages = {5416},
  doi = {10.1038/s41467-024-49862-1}
}

@article{hwang2013,
  title={Rounding of Au nano-crystals supported on sapphire at high temperatures},
  author={Hwang, Jae Sung and Noh, Do Young},
  journal={Journal of the Korean Physical Society},
  volume={62},
  number={1},
  pages={6--9},
  year={2013},
  publisher={Springer}
}

@article{kim2020,
  title={Laser-induced metastable mixed phase of AuNi nanoparticles: a coherent X-ray diffraction imaging study},
  author={Kim, Yoonhee and Kim, Chan and Ahn, Kangwoo and Choi, Jungwon and Lee, Su Yong and Kang, Hyon Chol and Noh, Do Young},
  journal={Synchrotron Radiation},
  volume={27},
  number={3},
  pages={725--729},
  year={2020},
  publisher={International Union of Crystallography}
}

@article{rheinheimer2020,
  title = {Equilibrium and Kinetic Shapes of Grains in Polycrystals},
  author = {Rheinheimer, Wolfgang and Blendell, John E. and Handwerker, Carol A.},
  year = 2020,
  month = jun,
  journal = {Acta Materialia},
  volume = {191},
  pages = {101--110},
  doi = {10.1016/j.actamat.2020.03.055}
}

@article{robinson2009,
  title = {Coherent {{X-ray}} Diffraction Imaging of Strain at the Nanoscale},
  author = {Robinson, Ian and Harder, Ross},
  year = 2009,
  month = apr,
  journal = {Nature Materials},
  volume = {8},
  number = {4},
  pages = {291--298},
  doi = {10.1038/nmat2400}
}

@article{saha2012,
  title = {Gold {{Nanoparticles}} in {{Chemical}} and {{Biological Sensing}}},
  author = {Saha, Krishnendu and Agasti, Sarit S. and Kim, Chaekyu and Li, Xiaoning and Rotello, Vincent M.},
  year = 2012,
  month = may,
  journal = {Chemical Reviews},
  volume = {112},
  number = {5},
  pages = {2739--2779},
  doi = {10.1021/cr2001178}
}

@article{si2021,
  title = {Gold Nanomaterials for Optical Biosensing and Bioimaging},
  author = {Si, Peng and Razmi, Nasrin and Nur, Omer and Solanki, Shipra and Pandey, Chandra Mouli and Gupta, Rajinder K. and Malhotra, Bansi D. and Willander, Magnus and De La Zerda, Adam},
  year = 2021,
  journal = {Nanoscale Advances},
  volume = {3},
  number = {10},
  pages = {2679--2698},
  doi = {10.1039/D0NA00961J}
}

@article{vicente2021,
  title = {Bragg {{Coherent Diffraction Imaging}} for {{{\emph{In Situ}}}} {{Studies}} in {{Electrocatalysis}}},
  author = {Vicente, Rafael A. and Neckel, Itamar T. and Sankaranarayanan, Subramanian K. R. S. and {Solla-Gullon}, Jos{\'e} and Fern{\'a}ndez, Pablo S.},
  year = 2021,
  month = apr,
  journal = {ACS Nano},
  volume = {15},
  number = {4},
  pages = {6129--6146},
  doi = {10.1021/acsnano.1c01080}
}

@article{wang2014,
  title = {Kinetics and {{Mechanisms}} of {{Aggregative Nanocrystal Growth}}},
  author = {Wang, Fudong and Richards, Vernal N. and Shields, Shawn P. and Buhro, William E.},
  year = 2014,
  month = jan,
  journal = {Chemistry of Materials},
  volume = {26},
  number = {1},
  pages = {5--21},
  doi = {10.1021/cm402139r}
}

@article{yang2022,
  title = {{\emph{In Situ}} {{Bragg}} Coherent {{X-ray}} Diffraction Imaging of Corrosion in a {{Co}}--{{Fe}} Alloy Microcrystal},
  author = {Yang, David and Phillips, Nicholas W. and Song, Kay and Barker, Clara and Harder, Ross J. and Cha, Wonsuk and Liu, Wenjun and Hofmann, Felix},
  year = 2022,
  journal = {CrystEngComm},
  volume = {24},
  number = {7},
  pages = {1334--1343},
  doi = {10.1039/D1CE01586A}
}

@article{zachman2022,
  title = {Measuring and Directing Charge Transfer in Heterogenous Catalysts},
  author = {Zachman, Michael J. and Fung, Victor and {Polo-Garzon}, Felipe and Cao, Shaohong and Moon, Jisue and Huang, Zhennan and Jiang, De-en and Wu, Zili and Chi, Miaofang},
  year = 2022,
  month = jun,
  journal = {Nature Communications},
  volume = {13},
  number = {1},
  pages = {3253},
  doi = {10.1038/s41467-022-30923-2}
}

@article{patterson1934,
  author  = {Patterson, A. L.},
  title   = {A {Fourier} Series Method for the Determination of the Components of Interatomic Distances in Crystals},
  journal = {Physical Review},
  year    = {1934},
  volume  = {46},
  number  = {5},
  pages   = {372--376},
  doi     = {10.1103/PhysRev.46.372},
  url     = {https://link.aps.org/doi/10.1103/PhysRev.46.372}
}

@article{mastropietro2017,
  title={Revealing crystalline domains in a mollusc shell single-crystalline prism},
  author={Mastropietro, Francesca and Godard, Pierre and Burghammer, Manfred and Chevallard, Corinne and Daillant, Jean and Duboisset, Julien and Allain, Marc and Guenoun, Patrick and Nouet, Julius and Chamard, Virginie},
  journal={Nature Materials},
  volume={16},
  number={9},
  pages={946--952},
  year={2017},
  publisher={Nature Publishing Group UK London}
}

@article{miao2012,
  title = {Coherent {{X-Ray Diffraction Imaging}}},
  author = {Miao, Jianwei and Sandberg, Richard L. and Song, Changyong},
  year = 2012,
  month = jan,
  journal = {IEEE Journal of Selected Topics in Quantum Electronics},
  volume = {18},
  number = {1},
  pages = {399--410},
  doi = {10.1109/JSTQE.2011.2157306}
}

@article{minkevich2008,
  title = {Applicability of an Iterative Inversion Algorithm to the Diffraction Patterns from Inhomogeneously Strained Crystals},
  author = {Minkevich, A. A. and Baumbach, T. and Gailhanou, M. and Thomas, O.},
  year = 2008,
  month = nov,
  journal = {Physical Review B},
  volume = {78},
  number = {17},
  pages = {174110},
  doi = {10.1103/PhysRevB.78.174110}
}

@article{ulvestad2016,
  title = {In {{Situ 3D Imaging}} of {{Catalysis Induced Strain}} in {{Gold Nanoparticles}}},
  author = {Ulvestad, Andrew and Sasikumar, Kiran and Kim, Jong Woo and Harder, Ross and Maxey, Evan and Clark, Jesse N. and Narayanan, Badri and Deshmukh, Sanket A. and Ferrier, Nicola and Mulvaney, Paul and Sankaranarayanan, Subramanian K. R. S. and Shpyrko, Oleg G.},
  year = 2016,
  month = aug,
  journal = {The Journal of Physical Chemistry Letters},
  volume = {7},
  number = {15},
  pages = {3008--3013},
  doi = {10.1021/acs.jpclett.6b01038}
}

@article{clark2012,
  title = {High-Resolution Three-Dimensional Partially Coherent Diffraction Imaging},
  author = {Clark, J.N. and Huang, X. and Harder, R. and Robinson, I.K.},
  year = 2012,
  month = aug,
  journal = {Nature Communications},
  volume = {3},
  number = {1},
  pages = {993},
  doi = {10.1038/ncomms1994}
}

@article{thompson2022,
  title = {{{LAMMPS}} - a Flexible Simulation Tool for Particle-Based Materials Modeling at the Atomic, Meso, and Continuum Scales},
  author = {Thompson, Aidan P. and Aktulga, H. Metin and Berger, Richard and Bolintineanu, Dan S. and Brown, W. Michael and Crozier, Paul S. and In 'T Veld, Pieter J. and Kohlmeyer, Axel and Moore, Stan G. and Nguyen, Trung Dac and Shan, Ray and Stevens, Mark J. and Tranchida, Julien and Trott, Christian and Plimpton, Steven J.},
  year = 2022,
  month = feb,
  journal = {Computer Physics Communications},
  volume = {271},
  pages = {108171},
  doi = {10.1016/j.cpc.2021.108171}
}

@article{ackland1987,
  title = {Simple {N}-body potentials for the noble metals and nickel},
  author = {Ackland, G. J. and Tichy, G. and Vitek, V. and Finnis, M. W.},
  year = {1987},
  journal = {Philosophical Magazine A},
  volume = {56},
  number = {6},
  pages = {735--756},
  doi = {10.1080/01418618708204485}
}

\end{document}